\documentclass[%
reprint,
amsmath,amssymb,
aps,
prd,
superscriptaddress,
]{revtex4-2}

\usepackage{ORCIDinREVTeX}
\usepackage{graphicx}
\usepackage{dcolumn}
\usepackage[caption=false]{subfig}
\usepackage{lmodern} 
\usepackage{courier} 
\usepackage[section]{placeins}
\usepackage{natbib}
\usepackage{bm}
\usepackage{hyperref}
\hypersetup{colorlinks=true, citecolor=cyan, urlcolor = blue, linkcolor = blue}

\usepackage{siunitx}
\usepackage{braket}

\usepackage{aas_macros}
\usepackage{color}

\DeclareUnicodeCharacter{2009}{\,}

\begin{document}

\title{Optimal Strategies for Gravitational-Wave Memory Detection with Noise Marginalization in Pulsar Timing Arrays}
\author{Rand Burnette}
\orcid{0009-0008-3649-0618}
\affiliation{Department of Physics, Oregon State University, Corvallis, OR 97331, USA}
\author{Xavier Siemens}
\orcid{0000-0002-7778-2990}
\affiliation{Department of Physics, Oregon State University, Corvallis, OR 97331, USA}
\affiliation{Center for Gravitation, Cosmology and Astrophysics, Department of Physics, University of Wisconsin-Milwaukee,\\ P.O. Box 413, Milwaukee, WI 53201, USA}

%


\date{\today}

\begin{abstract}
Current pulsar timing arrays could detect gravitational wave memory, which would provide a unique test of general relativity. In order to optimize our searches for gravitational wave memory in pulsar timing arrays, we integrate the optimal frequentist statistics for GW memory into the \texttt{enterprise} code suite. We additionally test a noise marginalized version of this statistic, and compare it's effectiveness to the simpler maximum posterior method. We perform these tests on simulated pulsar timing array with red and white noise content, as well as the NANOGrav $15$-year data set. We find no significant variations in our detection statistics between the noise-marginalized and the maximum-posterior methods for the simulated datasets, and only small variations in the NANOGrav $15$-year comparisons. We conclude that for realistic PTAs the noise-marginalization method could be useful for assigning a false alarm probability to Bayesian results, though the maximum posterior method is sufficient for identifying potentially detectable signals.
\end{abstract}

\maketitle


\section{\label{sec:intro} Introduction} 

Pulsar Timing Arrays (PTA) use precise timing of radio pulses from millisecond pulsars (MSPs) with ultra-stable rotational periods \cite{lorimer_binary_2008} to search for gravitational waves (GW). In recent years, we have found strong evidence for the presence of a gravitational wave background (GWB)~\cite{agazie_nanograv_2023-1, agazie_nanograv_2023-3, xu_searching_2023, PPTA_DR3_data, ppta_dr3_gwb, EPTA_DR2_data, epta_collaboration_and_inpta_collaboration_second_2023}. The most likely source for the GWB is a population of supermassive binary black holes~\cite{agazie_nanograv_2023,  Agazie_BHBconstraints_2023, Anisotropy_2023}, but it could also arise from beyond the standard model physics in the early universe~\cite{nanograv_15yr_newphysics, Afzal_NewPhysics_2023}. PTA data have also been searched for continuous waves \cite{ellis_optimal_2012, falxa_searching_2023, arzoumanian_nanograv_2023, antoniadis_second_2024}, GW memory \cite{12p5yr_NG_BWM, madison_pulsar_2017, sun_implementation_2023, Agazie_2025_15yrMem, ünal2025nonlineargravitationalwavememory}, and generic bursts \cite{becsy_bayesian_2021, deng_searching_2023, taylor2024fastwaveletbasissearch} as potential deterministic GW signals we could see beyond the GWB. 

Over the past two decades, the PTA community has devoted a great deal of effort to develop and refine gravitational wave detection techniques in PTAs. This has resulted in a suite of Bayesian and frequentist methods that ensure rigor and redundancy in GW detection and upper limit claims \cite{Littenberg_Bayes_2009, Lentati_Baysian_efficient, Anholm_Optimal, Demorest_2013, Ellis_2013, Becsy_QuickCW, ellis_optimal_2012, Chamberlin_2015, GWMem_optimal_2025, Taylor_CWbayesian}. Bayesian analyses are robust to the broad noise posteriors typical of PTA analyses, but can be computationally expensive. Frequentist analyses, on the other hand, are computationally efficient because we can analytically maximize the likelihood ratio over (at least) some of the signal parameters, and avoids the Bayesian marginalization of high dimensional integrals. For fixed values of the noise parameters, however, they can be biased, and a hybrid Bayesian-frequentist technique called noise-marginalization can be used to mitigate these biases~\cite{Vigeland_2018}.

The optimal statistic for the GWB was the testbed for developing noise marginalized techniques, as the most likely detectable GW signal in PTAs. The work done by \cite{Anholm_Optimal} to establish this statistic was adapted in \cite{Vigeland_2018} to introduce noise marginalization. In this work, it was shown that noise marginalization of the optimal statistic reduces the signal recovery bias due to covariance between the intrinsic red noise in the individual pulsars and the GWB. As a result, noise marginalization has become the standard for GWB frequentist work~\cite{Johnson_2024}.

As PTAs prepare to search for signals beyond the GWB, we are working to bring the frequentist methods used for various deterministic GW signals on par with the GWB optimal statistic. Very recently \cite{sardesai2025optimalstrategiescontinuouswave} developed a noise-marginalized version of the $\mathcal{F}_p$ statistic introduced in \cite{ellis_optimal_2012} for continuous wave (CW) searches. Surprisngly, unlike the results for the GWB optimal statistic, the implementation of noise marginalization into the $\mathcal{F}_p$ statistic resulted in no notable difference in signal significance recovery or bias.

Another potentially detectable signal in PTAs is GW memory, which is a nonlinear effect resulting in a constant shift in the space time metric due to violent GW-emitting events, such as mergers of super massive black hole binaries (SMBHBs) \cite{thorne_gravitational-wave_1992, christodoulou_nonlinear_1991, wiseman_christodoulous_1991,islo_prospects_2019}. As a memory wavefront passes Earth or an MSP, its effect is to redshift or blueshift the train of pulses traveling towards Earth. In PTAs we measure the times of arrival (TOAs) of these radio pulses, and compare them to our predicted model for when we expect these pulses to arrive. The difference between the predicted timing model and the TOAs yields the ``timing residuals" we use in our analyses. GW memory appears as a sudden change in the rotational frequency of the pulsar that results in a linear drift in our timing residuals which we will refer to as a ramp. 

In this paper, we study the robustness of frequentist techniques to search for GW memory. We start from the frequentist statistics developed for GW memory in \cite{GWMem_optimal_2025}, and build upon this work to include noise-marginalization as was done for the GWB and continuous waves in \cite{Vigeland_2018, sardesai2025optimalstrategiescontinuouswave}. We apply this method to the incoherent $\mathcal{F}_i$ statistic for GW memory, then compare the precision and efficacy of this with the standard method, which uses the maximum-posterior noise values to calculate the $\mathcal{F}_i$. We directly compare the value of the $\mathcal{F}_i$ statistic and false alarm probabilities (FAP) calculated across the time span of simulated and realistic data sets. We then check if there is any benefit from noise-marginalization in the detection and characterization of GW memory signals.

We begin in \autoref{sec:Methods} by reviewing the $\mathcal{F}$ statistics derivation for GW memory signals in PTA data. Then we explain how we marginalize over the red noise in our proposed method. In \autoref{sec:Results} we detail the simulated PTA data sets we generate, and compare both statistical methods effectiveness at recovering the correct detection distributions. In \autoref{sec:15yrData} we also perform similar analysis on the NANOGrav $15$-year data set in \autoref{sec:15yrData}, following the work done in \cite{Agazie_2025_15yrMem}. Finally, we present our conclusions about the effectiveness of both statistical methods in \autoref{sec:Conclusion}.

\section{\label{sec:Methods} Methods} 

We begin by describing the form of a GW memory signal in a PTA, and giving a brief overview of the form of the frequentist $F_c$ and $F_i$ statistics derived in \cite{GWMem_optimal_2025}. We then detail the noise-marginalization (NM) process, and how we compare it to the maximum-posterior (MP) method.

\subsection{\label{subsec:Signal Model} Signal Model}

We write the time-series for the residuals induced by a GW memory wavefront passing Earth at time $t_0$, from the direction $\hat{\Omega}$, with a polarization $\psi$, strain $h_0$, for a pulsar in the direction of the unit vector $\hat{p}$ as the vector
\begin{equation}
\delta \mathbf{t}_{mem} = a (h_0, \psi, \hat{p}, \hat{\Omega}) \mathbf{A}(t;t_0).
\label{equ:singlePulsarSig}
\end{equation}

Here, $a(h_0, \psi, \hat{p}, \hat{\Omega})$ accounts for the spatial response of the GW memory, in particular,
\begin{equation}
a (h_0, \psi, \hat{p}, \hat{\Omega}) = h_0\left[\cos{(2\psi)} F^+(\hat{p},\hat{\Omega}) + \sin{(2\psi)} F^\times(\hat{p},\hat{\Omega})\right].
\end{equation}
The $F^+$ and $F^\times$ terms are the antenna pattern functions for the earth-pulsar system,
\begin{align}
    F^+(\hat{p},\hat{\Omega}) & = \frac{1}{2} \frac{ (\hat{m}\cdot\hat{p})^2 - (\hat{n}\cdot\hat{p})^2}{1+ \hat{\Omega} \cdot \hat{p}} \\
    F^\times(\hat{p},\hat{\Omega}) & = \frac{1}{2} \frac{ (\hat{m}\cdot\hat{p})(\hat{n}\cdot\hat{p})}{1+ \hat{\Omega} \cdot \hat{p}},
\end{align}
where $\hat{m}$ and $\hat{n}$ are orthonormal vectors on the plane perpendicular to the propagation of the wave. They can be written as
\begin{align}
    &\hat{m}  = \sin{(\phi)}\hat{x} - \cos{(\phi)}\hat{y} \\
    &\hat{n}  = -\cos{(\theta)}\cos{(\phi)}\hat{x} - \cos{(\theta)}\sin{(\phi)}\hat{y} + \sin{(\theta)}\hat{z},
\end{align}
where $\phi$ and $\theta$ and the polar and azimuthal angles for the sky-location of the GW source. 

The right had side of Eq.~\ref{equ:singlePulsarSig}, also contains a time dependent piece $\mathbf{A}$, which is
\begin{equation}
\mathbf{A}(t;t_0) = (t-t_0)\Theta(t-t_0), 
\end{equation}
where $t_0$ is the time when the memory wavefront passes the Earth. 

Equation \ref{equ:singlePulsarSig} only contains the so-called "Earth Term." The full effect of GW memory should also include the "Pulsar term," which results from the wavefront passing the pulsar. However, here we consider only the Earth term because it is difficult to distinguish a single pulsar GW memory event from a transient noise effect. For more details on this see \cite{van_haasteren_gravitational-wave_2010, cordes_detecting_2012, Agazie_2025_15yrMem}.

The full time-series residuals for a single pulsar in a PTA are
\begin{equation}\label{eq:residuals}
    \delta \mathbf{t} =  \delta \mathbf{t}_{mem} + M \boldsymbol{\epsilon} + \mathbf{n} + F\mathbf{a}.
\end{equation}
In this equation, $M$ is the design matrix which contains the basis functions for the timing model, $\boldsymbol{\epsilon}$ are the uncertainties in the timing model parameters, and $\textbf{n}$ are the white noise uncertainties. The pulsar intrinsic red noise and common red noise (e.g. the GWB) are modeled with a Fourier basis $F$ and Fourier coefficients $\textbf{a}$.

We model each of the red noise processes using a power law \cite{phinney_practical_2001}
\begin{equation}\label{eq:red_noise_power}
    P_{RN}(f) =  A_{RN}^2 \left( \frac{f}{f_{1yr}} \right)^{-\gamma}.
\end{equation}
Here, $A_{RN}$ is the red noise amplitude at the reference frequency $f_{1yr}$, and $\gamma$ is the spectral index.

\subsection{\label{subsec:Fstats} The $\mathcal{F}_i$ statistic}

In this sub-section, we work through the derivation of the GW memory incoherent statistic $\mathcal{F}_i$. We follow formalism derived in \cite{GWMem_optimal_2025} for the frequentist statistical methods, while updating the likelihood definition to be compatible with the \texttt{enterprise} \cite{Johnson_2024} code base.

In general, the $\mathcal{F}$ statistics are obtained by maximizing the log-likelihood ratio \cite{Ellis_2013, Anholm_Optimal, GWMem_optimal_2025}. In a PTA, we will define the likelihood for a GW memory signal, following the convention outlined in \cite{Johnson_2024}
\begin{equation}\label{eq:mem_liklihood}
    p(\delta \mathbf{t} | \delta \mathbf{t}_{mem}) =  \frac{\exp[-\frac{1}{2} (\delta \mathbf{t} - \delta \mathbf{t}_{mem})^T C^{-1} (\delta \mathbf{t} - \delta \mathbf{t}_{mem})]}{\sqrt{\det(2 \pi C)}}.
\end{equation}
Here, the matrix $C$ is the timing model and red noise marginalized covariance matrix
\begin{equation}\label{eq:C}
    C = N + TBT^T.
\end{equation}
$N$ is the white noise covariance matrix
\begin{equation}\label{eq:N}
    N = \braket{\mathbf{nn}^T},
\end{equation}
$T$ contains the design matrix and Fourier basis functions
\begin{align}
    \label{eq:T}
    T = \begin{bmatrix}
        M \\
        F \\
    \end{bmatrix},
\end{align}
and $B$ is the covariance of the vector components of the timing model and Fourier coefficients
\begin{align}
    \label{eq:B}
    B = \begin{bmatrix}
        \braket{\boldsymbol{\epsilon} \boldsymbol{\epsilon}^T} & 0 \\
        0 & \braket{\mathbf{c} \mathbf{c}^T} \\
    \end{bmatrix}.
\end{align}
We take the inverse of this matrix using the Woodbury matrix identity to get
\begin{equation}\label{eq:Cinv}
    C^{-1} = N^{-1} - N^{-1} T (B^{-1} + T^TN^{-1}T)^{-1} T^T N^{-1}.
\end{equation}

We take the signal present likelihood, divided by the null likelihood, and take the log of that to get
\begin{equation}
\ln \Lambda = \ln \frac{p(\delta \mathbf{t} | \delta \mathbf{t}_{mem})}{p(\delta \mathbf{t} | 0)} = (\delta \mathbf{t} | \delta \mathbf{t}_{mem}) - \frac{1}{2} (\delta \mathbf{t}_{mem} | \delta \mathbf{t}_{mem}),
\end{equation}
with the inner product for two time domain vectors defined as 
\begin{equation}
(\mathbf{x} | \mathbf{y}) = \mathbf{x}^T C^{-1} \mathbf{y}.
\end{equation}

Maximizing this likelihood ratio will give us the highest likelihood GW memory signal given the data. We re-write this by defining the signal as a sum of amplitude and template vectors
\begin{equation}
\delta \mathbf{t}_{mem} = \sum_i a_i \mathbf{A}^i,
\label{equ:signaldecomp}
\end{equation}
which lets us write the log likelihood ratio as
\begin{equation}
\ln \Lambda = a_i (\delta \mathbf{t} | \mathbf{A}^i) - \frac{1}{2}a_i a_j (\mathbf{A}^i | \mathbf{A}^j).
\end{equation}

Here we define $\mathbf{N}^i = (\delta \mathbf{t} | \mathbf{A}^i)$ and $\mathbf{M}^{ij} = (\mathbf{A}^i | \mathbf{A}^j)$. Then we set the derivative of this equal to zero to maximize the likelihood ratio
\begin{equation}
\frac{\partial \ln \Lambda}{\partial a_j} = 0 = \mathbf{N}^{i} - a_i \mathbf{M}^{ij},
\end{equation}
which gives us both the maximum likelihood estimators for the "$a$" parameters 
\begin{equation}
a_i = \mathbf{M}_{ij}\mathbf{N}^{j},
\end{equation}
where $\mathbf{M}_{ij} = (\mathbf{M^{-1}})^{ij}$, as well as the corresponding max likelihood ratio
\begin{equation}
\ln \Lambda = \frac{1}{2}\mathbf{N}^{i}\mathbf{M}_{ij}\mathbf{N}^{j} \equiv \mathcal{F}.
\end{equation}
The average over many noise realizations of the $2\mathcal{F}$ statistic follows a $\chi^2$ distribution, where the decomposition of the signal as mentioned above can be done coherently or incoherently across pulsars. 

In this paper we focus on the incoherent $\mathcal{F}_i$ statistic for GW memory. There is a coherent version of this statistic, $F_c$, that we do not apply the NM method too because of the prohibitive computational cost. As the coherent statistic depends on the sky-location of the signal source, we need to marginalize over the sky locations as well as the red noise parameters. This results in an increased computational time, from hours to days, and renders the NM $F_c$ statistic unusable, with a higher computational cost than traditional Bayesian methods \cite{Agazie_2025_15yrMem}. For similar reasons the authors of~\cite{sardesai2025optimalstrategiescontinuouswave} abandoned the coherent NM statistic for continuous waves.

In the incoherent case, \autoref{equ:signaldecomp} is a sum over the signals of individual pulsars we defined in \autoref{equ:singlePulsarSig}, so $\delta \mathbf{t}_{mem} = \sum_\alpha^N a_\alpha \mathbf{A}^\alpha$ where $N$ is the number of pulsars.  We can write the value of the statistic $\mathcal{F}_\alpha$ for each pulsar as
\begin{equation}
\mathcal{F}_\alpha = \frac{1}{2}\mathbf{N}^2\mathbf{M}^{-1},
\end{equation}
so total $\mathcal{F}_i = \sum^N_\alpha  \mathcal{F}_\alpha$. This means $2\mathcal{F}_i$ statistic, follows a $\chi^2$ distribution with $N$ degrees of freedom, and, if a signal is present, with a non-centrality parameter $\rho^2 = (\tilde{\delta \mathbf{t}}_{mem} | \tilde{\delta \mathbf{t}}_{mem})$
\begin{equation}
\braket{2\mathcal{F}_i} = N + \rho^2,
\end{equation}
see~\cite{sun_implementation_2023} for more details.

\subsection{\label{subsec:FalseAlarmMath} False Alarm Probability}

In order to use the $\mathcal{F}_i$ statistic as a robust detection statistic, we  calculate false alarm probabilities (FAPs) for all of our simulations. The FAP is the chance that the value obtained for the $\mathcal{F}_i$ statistic could have been created by noise alone, and is calculated by comparing the recovered $\mathcal{F}_i$ value to the null distribution of the statistic. As the $\braket{2\mathcal{F}_i}$ follows a $\chi^2$ distribution with $N$ degrees of freedom, we have an analytic form of the null distribution. This follows the same structure as other deterministic signal $\mathcal{F}$-statistics \cite{sardesai2025optimalstrategiescontinuouswave, Jaranowski_FAP_background}, and we write as
\begin{equation}
p_0(\mathcal{F}_i) = \frac{\mathcal{F}_i^{\frac{N}{2}-1}}{(\frac{N}{2}-1)!}\exp(\mathcal{F}_i).
\label{eq:nulldist}
\end{equation}
The FAP for a $\mathcal{F}_i$ value $\mathcal{F}_0$ is found by taking the integral over the null distribution from this value to infinity, which represents the portion of the null distribution that could produce a value of $\mathcal{F}_0$ or larger
\begin{equation}
P^T_F(\mathcal{F}_0) = \int^\infty_{\mathcal{F}_0}p_0(\mathcal{F}_i) d\mathcal{F}_i.
\end{equation}

\subsection{\label{subsec:Methods} Maximum-Posterior and Noise-Marginalized Methods}

In order to calculate the $\mathcal{F}_i$ statistic for a PTA data set, we need a set of noise model parameters to construct our covariance matrix with. Here, we will discuss two methods that can be used when deciding the noise model parameters.

All modern PTA data sets contain some amount of red noise and white noise. The white noise values for each pulsar can be determined by a Bayesian analysis, such as was done for the NANOGrav 15-year data set \cite{agazie_nanograv_2023-3}. However, there are two established methods for determining the red noise parameter values. Both start with performing another Bayesian search over the intrinsic RN and CURN parameters, so that you obtain a chain of parameter values that you can construct parameter posteriors from. 

One method is to find the maximum likelihood value obtained in the Bayesian chain, and set the red noise parameters to the corresponding values. We call this the \textbf{maximum-posterior (MP)} method. The second method involves performing many random draws of the parameters from the chain, and calculating the $\mathcal{F}_i$ values for each set of red noise parameters to obtain a distribution for the $\mathcal{F}_i$ statistic over the red noise posterior values. This we refer to as the \textbf{noise-marginalized (NM)} method. It should be noted that the NM method takes more than an order of magnitude longer to run, as all randomly sampled red noise parameters need a new calculation of the statistic using a different covariance matrix in the scalar products.

In work done with the GWB optimal statistic in \cite{Vigeland_2018}, it was shown that when the signal being searched for is highly covariant with other red noise processes, this NM method is more robust. However, \cite{sardesai2025optimalstrategiescontinuouswave} found that for continuous GWs the MP method is sufficient. Therefore, one of the goals of this work is to determine whether the more computationally intensive NM method is needed for GW memory detection statistics.

\section{\label{sec:Results} PTA Statistic Analysis} 

In this section, we describe the simulated and realistic data sets we use to test the $\mathcal{F}_i$ statistic. We then detail the $\mathcal{F}_i$ and FAP results for these data sets calculated using both the MP and NM methods.

All simulated data sets in this section consist of a "Theorist's PTA," which has $20$ pulsars distributed randomly across the sky with $20$ years of data. Each pulsar has a gaussian white noise process with $\sigma = \SI{100}{ns}$, and shares a CURN process with $\log_{10}A=-15$ (unless specified) and $\gamma=4.33$. All simulations also contain a GW memory signal with a starting epoch of $t_0=\SI{58109}{MJD}$, with a variety of strain amplitudes.

\begin{figure}[htbp]
     \centering
     \begin{subfloat}
         \centering
         \includegraphics[width=1\columnwidth]{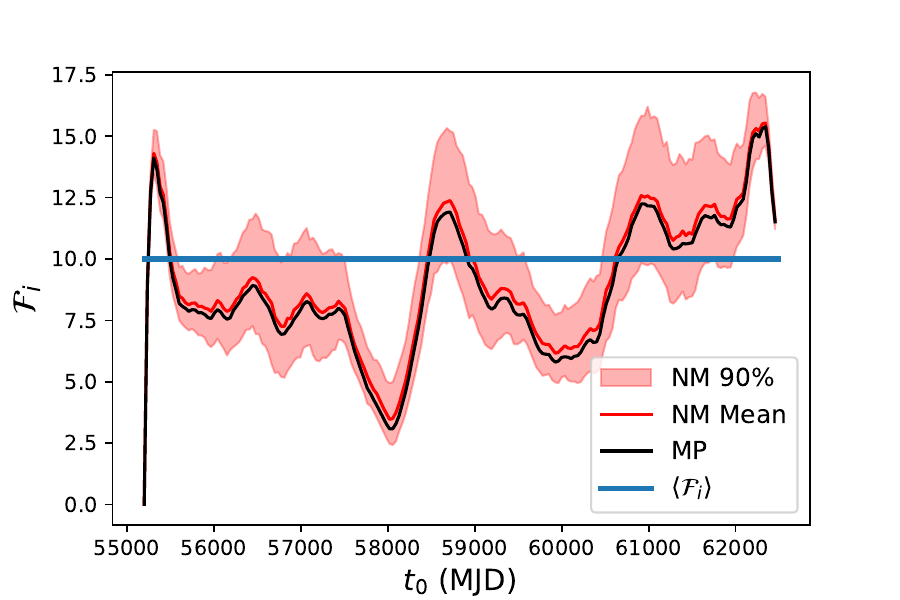}
     \end{subfloat}
     \begin{subfloat}
         \centering
         \includegraphics[width=1\columnwidth]{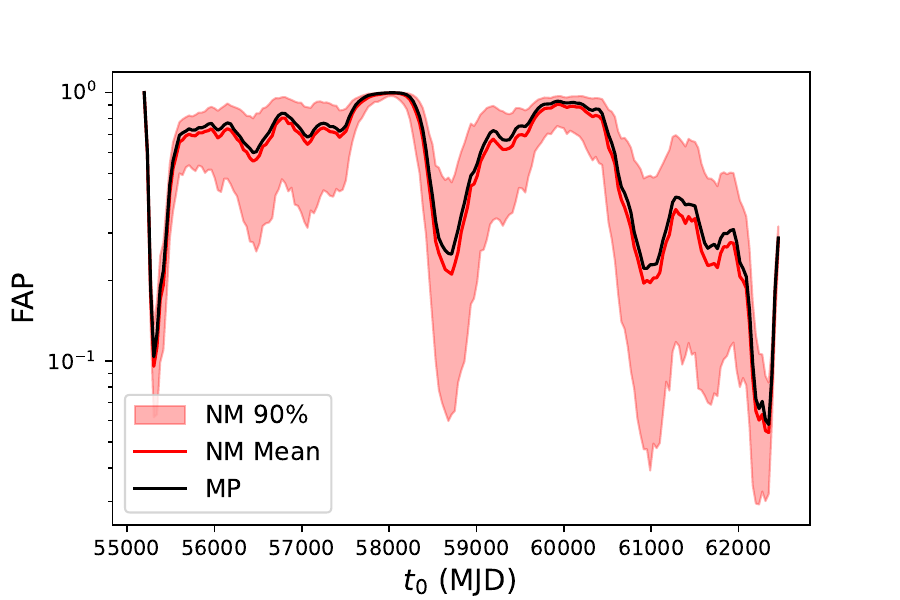}
     \end{subfloat}
     \caption{These figures show the results of analysis done on simulated data sets containing WN of $\sigma = 100ns$, a CURN with $\gamma=4.33$ and $A=10^{-15}$, as well as a low amplitude memory signal with $h=10^{-15}$. \textbf{Top:} the values of the $\mathcal{F}_i$ statistic evaluated using the MP parameter values and the NM method over a range of $200$ memory $t_0$ values. Over plotted in blue is the average value of the null distribution for $\mathcal{F}_i$, which is half the number of pulsars. \textbf{Bottom:} The FAP distributions calculated from the $\mathcal{F}_i$ results.}
     \label{fig:LowSigPlots}
\end{figure}

\subsection{\label{subsec:Simulations} Varied Signal Strength Simulations}

We show results for three simulated data sets with different GW memory signal amplitudes. A "low amplitude" simulation with $\log_{10}h_0=-15$, a "mid amplitude" with $\log_{10}h_0= -13.35$, and a "high amplitude" simulation with $\log_{10}h_0= -13$. We run the $\mathcal{F}_i$ analysis on all three data sets, using both red noise parameter estimation methods, across a grid of 200 MJD values for the GW memory ramp $t_0$. All plots include the mean $\mathcal{F}_i$ or FAP value of the NM method, along with a $90\%$ credible region showing the distribution plotted in red. They also have the value calculated from the MP method over plotted in black.

\begin{figure}[hbp]
     \centering
     \begin{subfloat}
         \centering
         \includegraphics[width=1\columnwidth]{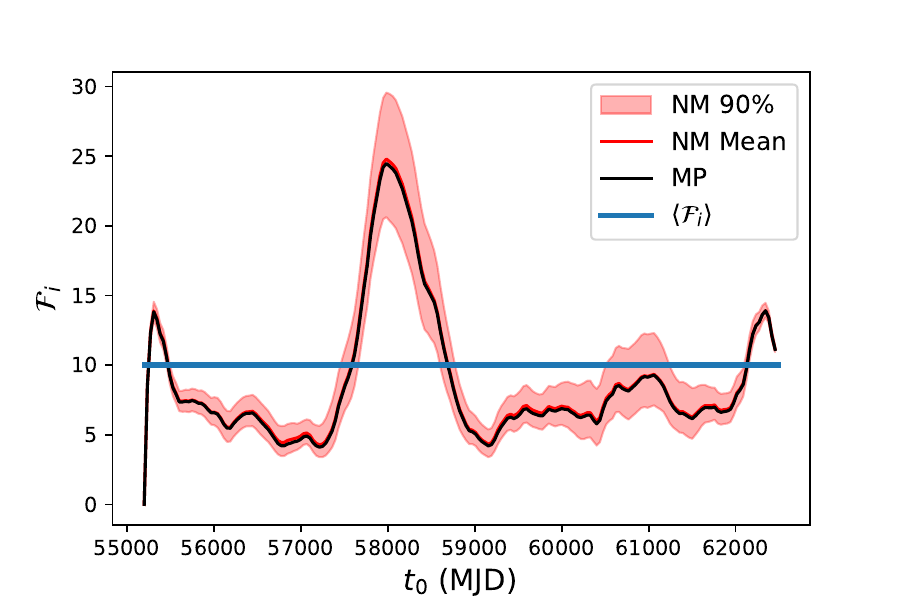}
     \end{subfloat}
     \begin{subfloat}
         \centering
         \includegraphics[width=1\columnwidth]{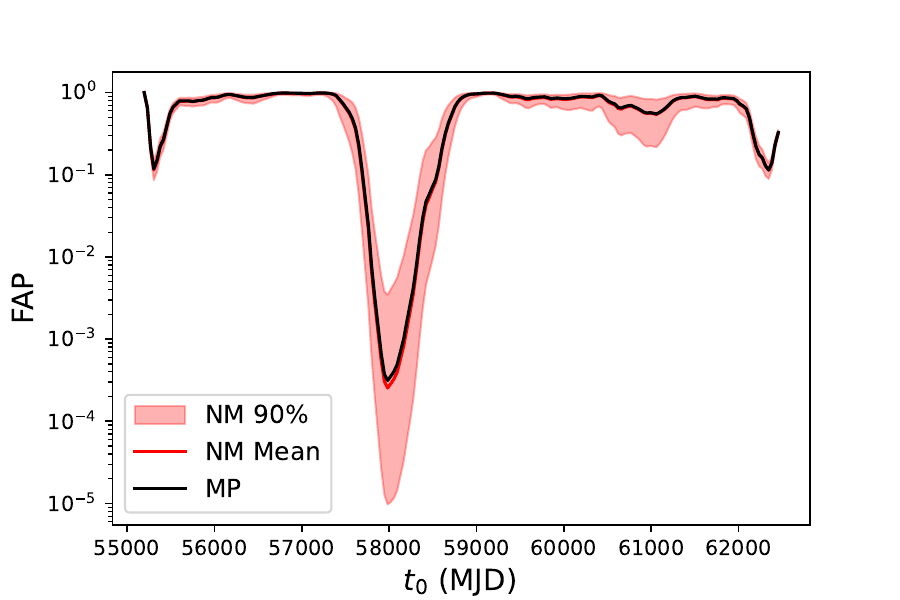}
     \end{subfloat}
     \caption{These figures show the results of analysis done on simulated data sets containing WN of $\sigma = 100ns$, a CURN with $\gamma=4.33$ and $A=10^{-15}$, as well as a medium amplitude memory signal with $h=10^{-13.35}$. \textbf{Top:} the values of the $\mathcal{F}_i$ statistic evaluated using the MP parameter values and the NM method over a range of $200$ memory $t_0$ values. Over plotted in blue is the average value of the null distribution for $\mathcal{F}_i$, which is half the number of pulsars. \textbf{Bottom:} The FAP distributions calculated from the $\mathcal{F}_i$ results.}
     \label{fig:MidSigPlots}
\end{figure}

The low amplitude simulation results are plotted in \autoref{fig:LowSigPlots}. The distribution of $\mathcal{F}_i$ and FAP values across the $t_0$ range show no significant excursions, which we expect given the strength of GW memory present. The highest peaks, landing at FAP values of around $1\%$, appear at the edges of the data set. This reflects a known lack of sensitivity PTAs have to GW memory near the edges due to timing model interference, see \cite{Agazie_2025_15yrMem} for a more detailed discussion. 

In \autoref{fig:MidSigPlots} we show the results for the mid amplitude signal. In that more sensitive central region there is now a peak where there had been a valley around the expected epoch $t_0 = \SI{58109}{MJD}$. The FAP values show that there is stronger evidence for a GW memory event, but as the mean NM FAP and MP FAP are still around the $0.1\%$ level, it is not enough to claim a detection.

\begin{figure}[htbp]
     \centering
     \begin{subfloat}
         \centering
         \includegraphics[width=1\columnwidth]{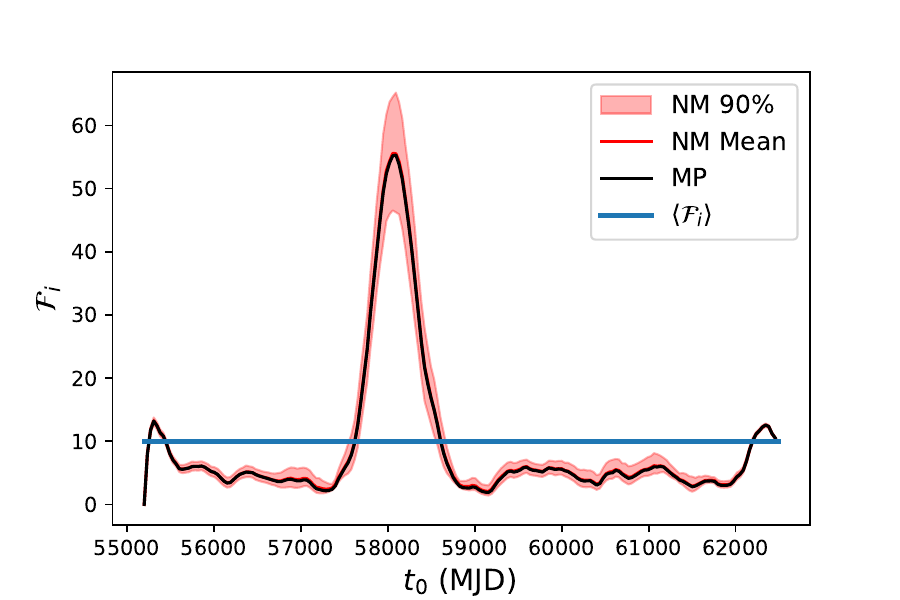}
     \end{subfloat}
     \begin{subfloat}
         \centering
         \includegraphics[width=1\columnwidth]{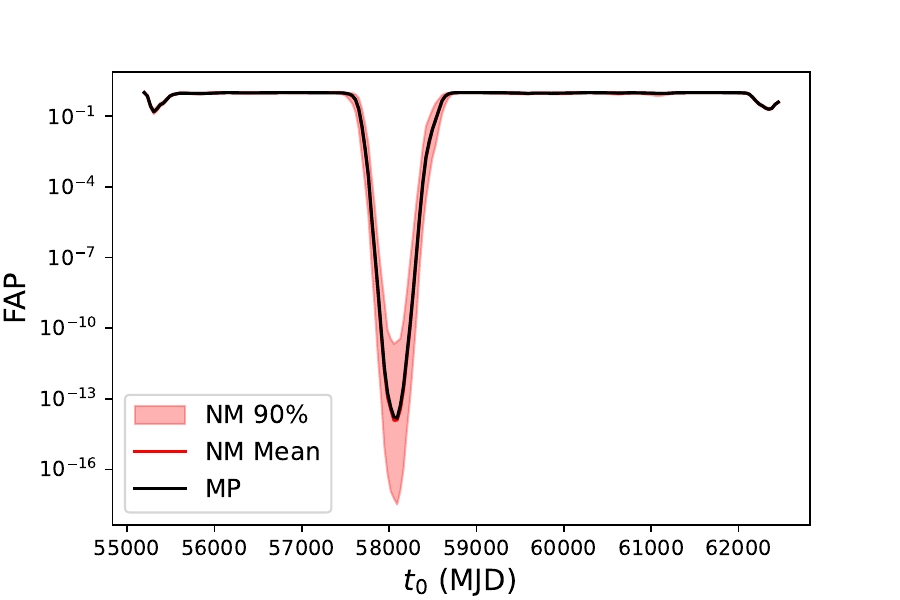}
     \end{subfloat}
     \caption{These figures show the results of analysis done on simulated data sets containing WN of $\sigma = 100ns$, a CURN with $\gamma=4.33$ and $A=10^{-15}$, as well as a Large amplitude memory signal with $h=10^{-13}$. \textbf{Top:} the values of the $\mathcal{F}_i$ statistic evaluated using the MP parameter values and the NM method over a range of $200$ memory $t_0$ values. Over plotted in blue is the average value of the null distribution for $\mathcal{F}_i$, which is half the number of pulsars. \textbf{Bottom:} The FAP distributions calculated from the $\mathcal{F}_i$ results.}
     \label{fig:LargeSigPlots}
\end{figure}

The results of the large GW memory injection in \autoref{fig:LargeSigPlots} show an unmistakable peak present. Centered around the correct $t_0$ value, we get FAP value around $10^{-13}$ for both detection methods. We can better understand why we obtain this sudden increase in detection probability when we consider the null distribution for the $\mathcal{F}_i$ we defined in \autoref{eq:nulldist}. The null distribution is centered around $\frac{N}{2}$, which is half the number of pulsars in the PTA for the $\mathcal{F}_i$ statistic. This means values of 5 to 15 are well within the expected noise distribution for this data set, but $\mathcal{F}_i = 60$ is well in the tail of what could be produced only using the noise present.

\subsection{\label{subsec:Simulations} Varied Red Noise Content Simulations}

In previous work, we have seen that the level of common red noise present in the dataset can have a significant impact on the significance of a GW memory detection \cite{Agazie_2025_15yrMem}. To test this with the $\mathcal{F}_i$ methods, we perform a set of tests on three datasets with a small GW memory signal of $\log_{10}h_0= -13.5$. We then reduce the level of red noise present in each simulation from our standard level $\log_{10}A=-15$, to a reduced significance $\log_{10}A=-15.5$, and finally to effectively no red noise content with $\log_{10}A=-18$. 

\begin{figure}[htbp]
     \centering
     \begin{subfloat}
         \centering
         \includegraphics[width=1\columnwidth]{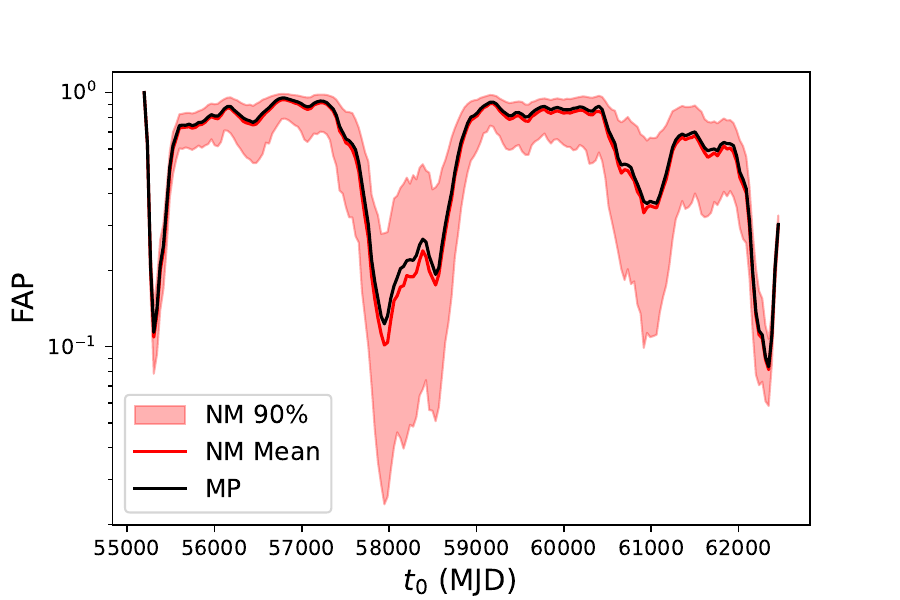}
     \end{subfloat}
     \begin{subfloat}
         \centering
         \includegraphics[width=1\columnwidth]{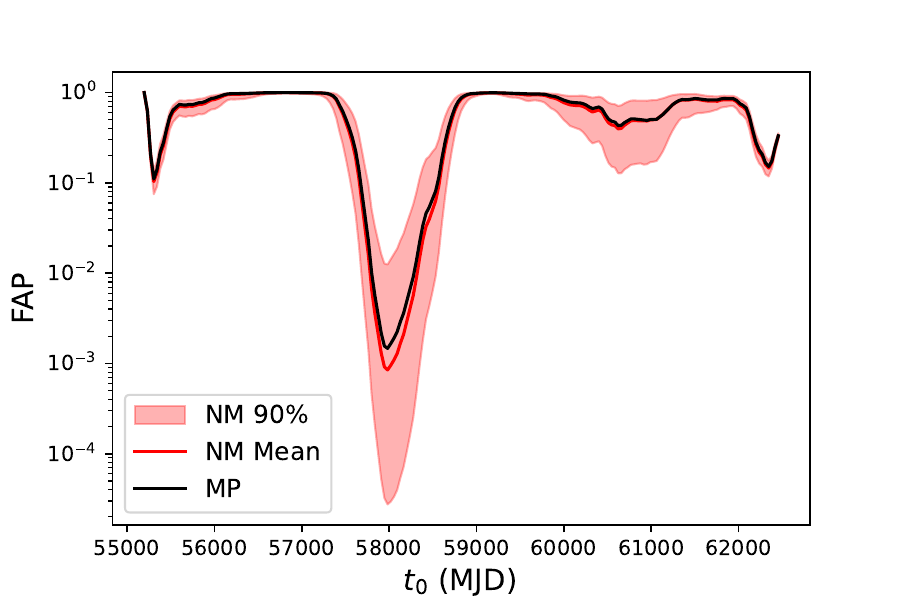}
     \end{subfloat}
     \begin{subfloat}
         \centering
         \includegraphics[width=1\columnwidth]{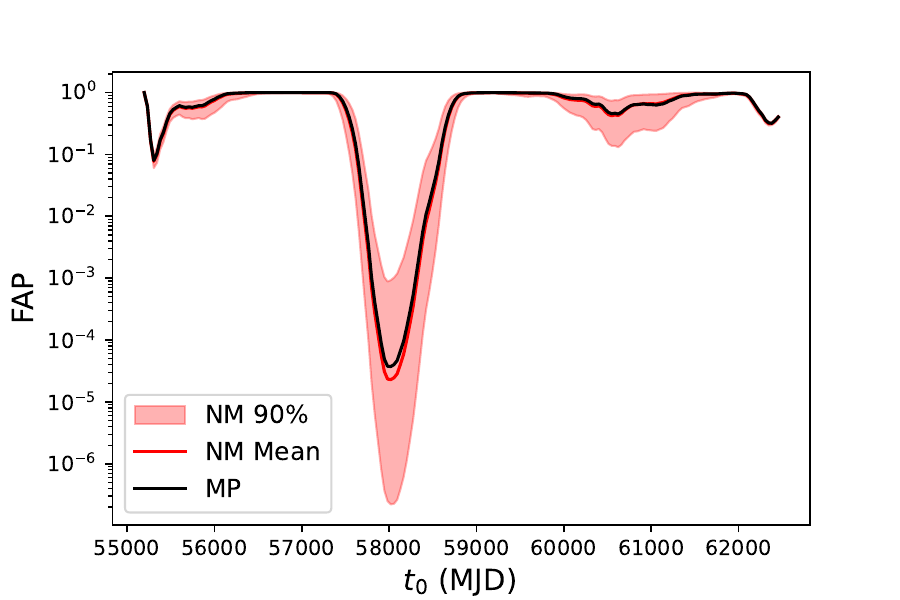}
     \end{subfloat}
     \caption{These figures show FAP plots for three simulations containing a WN process of $\sigma = 100ns$, as well as a memory signal with $h=10^{-13.5}$. From top to bottom, the data sets are distinguished by their red noise content of $\log_{10}A=-15$, $\log_{10}A=-15.5$, and finally  $\log_{10}A=-18$.}
     \label{fig:CurnVarySigPlots}
\end{figure}

The FAP results for these three datasets are shown in \autoref{fig:CurnVarySigPlots}. Ordered in descending red noise amplitude order from top to bottom, we can see that a decrease in red noise content can change a low significance GW memory signal into a detection. 

Across all of these simulated results we see a very consistent result when comparing the two methods of estimating the red noise parameters. The MP method values almost completely overlap with the mean values of the NM method, and they all fall within the $90$ percentile region. In each simulated data set, the resulting inference possible from the $\mathcal{F}_i$ statistic does not change when using the MP or NM method.

\begin{figure}[htbp]
    \centering
    \begin{subfloat}
         \centering
         \includegraphics[width=1\columnwidth]{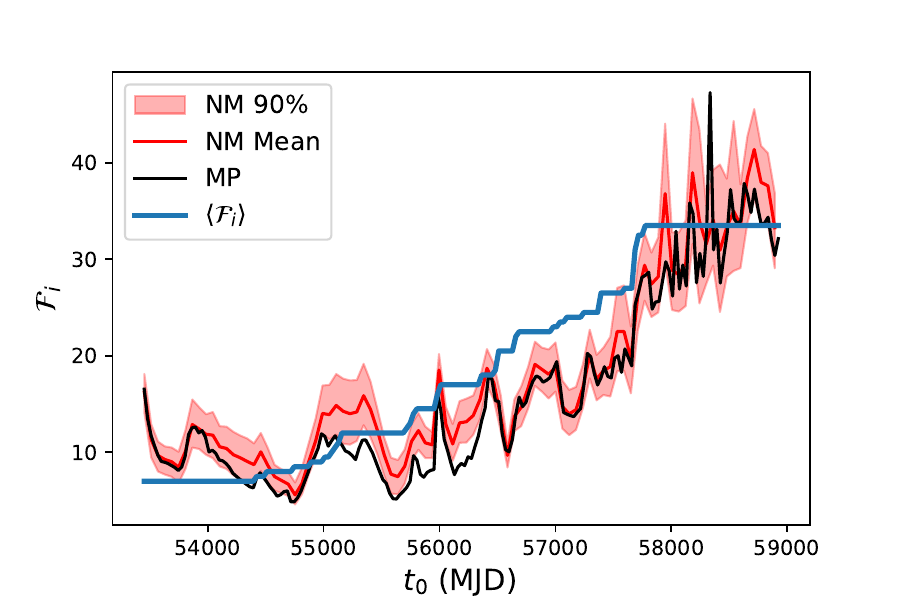}
    \end{subfloat}
    \begin{subfloat}
         \centering
         \includegraphics[width=1\columnwidth]{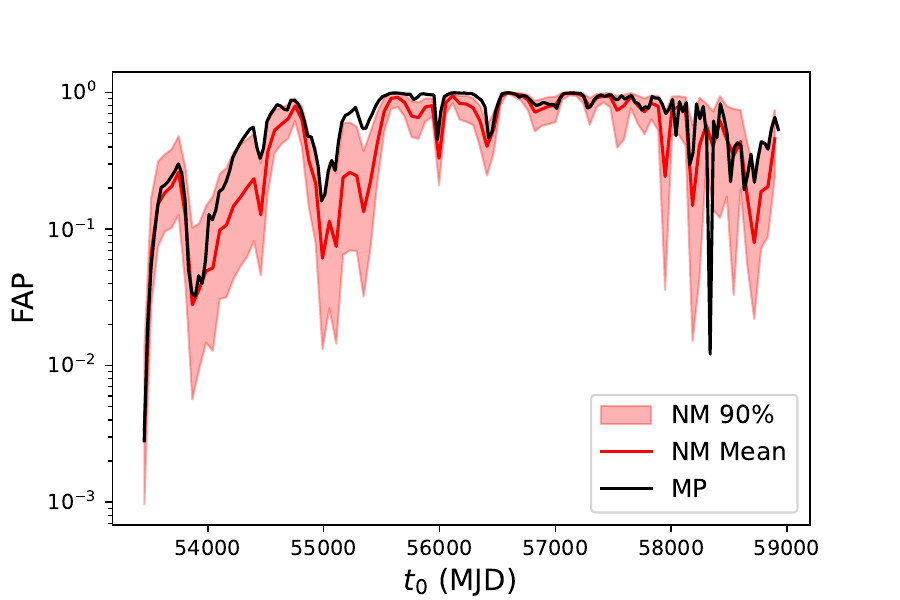}
    \end{subfloat}
    \caption{\textbf{Top:} the $\mathcal{F}_i$ results for the NANOGrav 15yr data set. Plotted are the MP results with the NM distributions for each $t_0$ bin. Over plotted in blue is the average value of the null distribution for $\mathcal{F}_i$. That value is half the number of degrees of freedom in the $2\mathcal{F}_i$ statistic, which is the number of pulsars. As this data set does not have evenly sampled pulsars, the number of pulsars that can observe a GW memory event change over the time period, and so does the $\mathcal{F}_i$ null distribution. \textbf{Bottom:} The False Alarm Probabilities, calculated using the $\mathcal{F}_i$ values and taking into account the varying number of pulsars over the data set time span.}
    \label{fig:15yrSNRplots}
\end{figure}

\subsection{\label{sec:15yrData} A Realistic Data Set: the NANOGrav 15-year data} 

In addition to the simulated results, we compare the effectiveness of the MP and NM $\mathcal{F}_i$ methods and compare them to the results found in the  Bayesian search in the NANOGrav 15-year data set for GW memory \cite{Agazie_2025_15yrMem}. In that paper, we found no significant evidence for GW memory, with some minor spikes in significance near the edges of the data sets. We performed a set of MP and NM runs on this data set and show both $\mathcal{F}_i$ as well as FAP results using these methods in \autoref{fig:15yrSNRplots}. Both statistics were generated using posteriors from a Bayesian run searching for uncorrelated red noise (CURN) in the 15yr data set. Starting with $\mathcal{F}_i$, we see that over the full span of the 15-year data set timing baseline, the largest spike occurs in the MP results near the end of the data set, just after $58000$ MJD. This is near where the largest peak occurred in the posteriors of the 15-year GW Memory Bayesian search \cite{Agazie_2025_15yrMem}. However, when we are considering detection thresholds for the GW memory signal, we need to calculate the FAP values. 

In order to calculate the FAPs in a real data set, we need to revisit the null distribution of our statistic. As we showed in \autoref{eq:nulldist}, the null distribution depends on the number of pulsars. In the 15-year data set, pulsars have very different timing baselines, and so different numbers of pulsars will see the point where the GW memory ramp begins ($t_0$). This means that the null distribution changes over the range of $t_0$, which we have done for the FAP plots shown in \autoref{fig:15yrSNRplots}. We have also cut off $200$ days from the start of the plot because there were fewer than $10$ pulsars that could see burst from that epoch, and this inflates the FAPs we see due to known edge effects \cite{Agazie_2025_15yrMem}. We see no points that are below $\sim 1\%$ FAP. 
\begin{figure}[htbp]
    \centering
    \includegraphics[width=1\columnwidth]{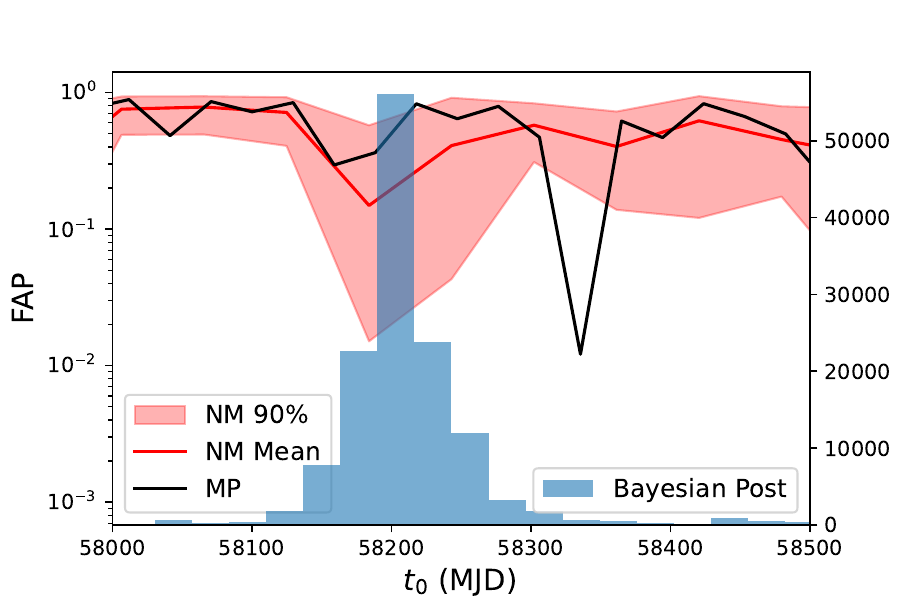}
    \caption{The False Alarm Probabilities for the NANOGrav 15yr data set plotted with the Bayesian posterior distribution for $t_0$ found in \cite{Agazie_2025_15yrMem}.}
    \label{fig:15yrFAPBayes}
\end{figure}

An interesting observation can be made if we compare these FAP results to the Bayesian posterior results in \cite{Agazie_2025_15yrMem}, see \autoref{fig:15yrFAPBayes}. If we over-plot the posterior for the burst epoch $t_0$ in the Bayesian result with the FAPs, and zoom into the most significant peak, we can see that it does not quite align with the MP peak. However, it overlaps one of the NM peaks. This is to be expected since the NM method is the better frequentist parallel to the Bayesian analysis.

We also compare the effectiveness of the MP versus NM methods. We see that across the majority of the timing baseline, both methods overlap. In the spots where the two methods do not quite align, there is still no significant difference in the conclusion that there is a non-detection, even though the NM results do tend towards lower FAPs.

\section{\label{sec:Conclusion} Conclusion} 

Our goal was to determine the optimal method for utilizing the $\mathcal{F}_i$ statistic as a detection statistic for GW memory in realistic PTA data sets. We have seen that across all our simulated data sets, with different GW memory significances and different red noise contents, the MP and NM methods produce equivalent results. They also both produce FAPs that effectively indicate the level of confidence we expected for signal detection. This gives us both a baseline for how we would expect a GW memory signal to appear in future PTA analysis, and indicates that the MP method is sufficient as a detection statistic without noise marginalization.

In addition to these tests, we performed the analysis on the NANOGrav $15$-year data set to test for the consistency of our results are with the previous published Bayesian results. We find that there is more variation between methods in real datasets, though both methods show non-detection results that are consistent with the Bayesian findings. This suggests that the MP method is sufficient for GW memory detection analysis using the $\mathcal{F}_i$. 

On closer inspection of the burst epoch that was preferred by the Bayesian results, we see that the NM results show more significance at that time than the MP method. Though for most preliminary results the MP method seems sufficient, the NM method seems to still more closely align with the Bayesian results, and could be useful for assigning FAP to such Bayesian peaks in future analyses. 

\section{\label{sec:Acknowledgements} Acknowledgments} 

We would like to thank Jerry Sun for the development of the original $\mathcal{F}_i$ statistic code that we have built off of, as well as Caner Unal for helping to review the paper. The NANOGrav collaboration receives support from National Science Foundation (NSF) Physics Frontiers Center award 2020265.

\pagebreak

\bibliography{NG_15_BWM_bib}{}
\bibliographystyle{unsrt_et_al_15-3author_notitle}

\end{document}